\documentclass[journal]{IEEEtran}
\usepackage{cite}
\ifCLASSINFOpdf
\usepackage[pdftex]{graphicx}
\else
\fi
\begin{document}

\title{Room temperature exchange bias in BiFeO$_3$ / Co-Fe bilayers}%
\author{
\IEEEauthorblockN{Christian~Sterwerf$^1$,~Markus~Meinert$^1$,~Elke~Arenholz,$^2$~Jan-Michael~Schmalhorst$^1$,~and~G\"unter~Reiss$^1$}\\[2mm]
\IEEEauthorblockA{\small{$^1$ Center for Spinelectronic Materials and Devices, Physics Department, Bielefeld University, Germany}}\\[1mm]
\IEEEauthorblockA{\small{$^2$ Advanced Light Source, Lawrence Berkeley National Laboratory, California 94720, USA}}
}
\maketitle
\date{\today}

\begin{abstract}
\boldmath
Thin highly epitaxial BiFeO$_3$ films were prepared on SrTiO$_3$ ($100$) substrates by reactive magnetron co-sputtering. Detailed MOKE measurements on BiFeO$_3$/Co-Fe bilayers were performed to investigate the exchange bias as a function of the films thicknesses and Co-Fe stoichiometries. We found a maximum exchange bias of H$_{\mathrm{eb}}=92$\,Oe and a coercive field of H$_{\mathrm{c}}=89$\,Oe for a $12.5$\,nm thick BiFeO$_3$ film with a $2$\,nm thick Co layer. The unidirectional anisotropy is clearly visible in in-plane rotational MOKE measurements. AMR measurements reveal a strongly increasing coercivity with decreasing temperature, but no significant change in the exchange bias field.%
\end{abstract}


\section{Introduction}
Multiferroic materials are materials that simultaneously exhibit spontaneous electric and magnetic ordering \cite{Eerenstein:2006km}. BiFeO$_3$ has large ferroelectric polarization and antiferromagnetic ordering. It has recently attracted a lot of interest because of its suitability for room temperature devices due to a high ferroelectric transition temperature of T$_{\mathrm{C}}=1100$\,K as well as a high N\'{e}el temperature of T$_{\mathrm{N}}=640$\,K \cite{Kiselev,Teague:1970tk}. Possible spintronic applications of the multiferroicity include control of the magnetization direction of an exchange coupled ferromagnetic layer with an electric field. BiFeO$_3$ has a G-type antiferromagnetic ordering in which the Fe magnetic moments are parallel within the ($111$) planes and alternate between adjacent planes. Thin BiFeO$_3$ films on SrTiO$_3$ ($100$) substrates crystallize in a tetragonal structure similar to the cubic perovskite structure with space group \textit{P4mm} and an in-plane lattice parameter of $a=3.935$\,\AA{} with a small tetragonal distortion \cite{Wang:2003ca}. A small displacement of the atoms within the cell induces the spontaneous electric polarization. In contrast, BiFeO$_3$ crystallizes in R3c structure in the bulk.

In this work we present detailed investigations of the exchange bias effect between antiferromagnetic BiFeO$_3$ films and different ferromagnetic layers as a function of both layer thicknesses.

\section{Preparation and Characterization Techniques}
Thin BiFeO$_3$ films were deposited using a UHV co-sputtering system with a base pressure of $1\cdot10^{-9}$\,mbar. The films were grown by dc- and rf- magnetron sputtering from elemental Bi and Fe targets onto SrTiO$_3$ ($100$) substrates. SrTiO$_3$ ($100$) substrates allow for a coherent and epitaxial growth of the BiFeO$_3$ films. The lattice parameter of the SrTiO$_3$ ($100$) substrates is $3.91$\,\AA{}, while the pseudo-cubic in-plane lattice parameter for BiFeO$_3$ is $3.935$\,\AA{}, resulting in a lattice mismatch of less than $0.7$\,\% \cite{Wang:2003ca}. The oxygen partial pressure and deposition temperatures were varied in order to optimize the growth conditions. The correct sputtering powers for Bi and Fe were determined according to x-ray fluorescence spectroscopy. In order to investigate the exchange bias of the BiFeO$_3$ films additional ferromagnetic Co-Fe layers were deposited on top of the BiFeO$_3$. To achieve pinning between both layers the ferromagnet was deposited at a substrate temperature of $175^\circ $C. Higher deposition temperatures for the Co-Fe layer led to excessive interdiffusion. Afterwards the bilayer was field cooled in a magnetic field of $150$\,Oe in $[100]$ direction of the BiFeO$_3$. Finally, a $2$\,nm thick MgO capping layer was deposited.

X-ray absorption spectroscopy (XAS), x-ray magnetic circular dichroism (XMCD) and x-ray linear dichroism (XLD) measurements were performed at BL 4.0.2 of the Advanced Light Source in Berkeley, CA, USA. The measurements were done at room temperature in surface-sensitive total electron yield (TEY) \cite{Idzerda:1994vq}. A magnetic field of $0.5$\,T was applied parallel to the beam. The linear (circular) polarization degree was $100$\% ($90$\%). The x-ray angle of incidence to the sample surface was $30^\circ$ for all measurements.
Measurements of the magneto-optic Kerr effect (MOKE) were performed at room-temperature in a system with a laser wavelength of $650$\,nm. For in-plane anisotropy measurements the sample was mounted on a rotating sample stage.
To measure the temperature dependence of the exchange bias, a $12.5$\,nm thick BiFeO$_3$ film with a $2.5$\,nm Co layer was patterned into a Hall-bar geometry with a width of $183\,\mu$m and a length of $622\,\mu$m and the anisotropic magnetoresistance (AMR) was recorded. These measurements were performed in a closed cycle He-cryostat in a temperature range from $20$\,K to $300$\,K.

\section{Crystallographic Properties}
\begin{figure}[!t]%
\centering%
\includegraphics[width=\linewidth]{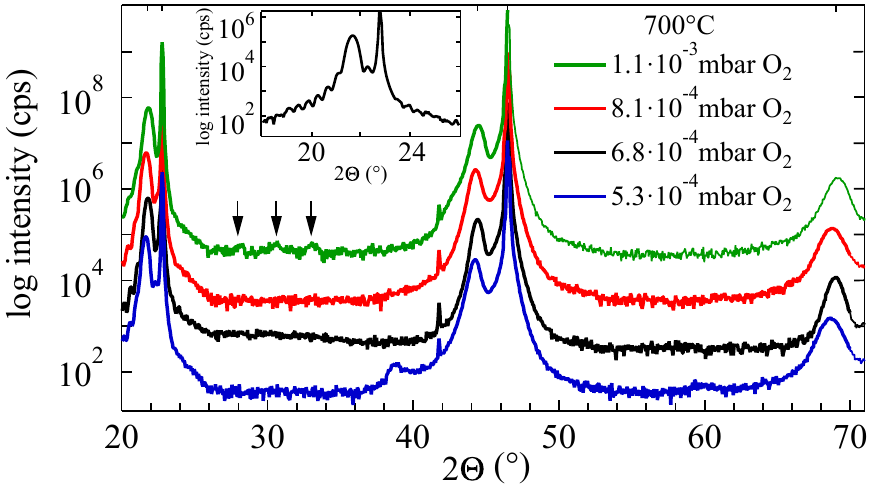}%
\caption{XRD patterns for BiFeO$_3$ films grown at a substrate temperature of $700^\circ$C with various oxygen partial pressures. The inset shows a magnification of the BiFeO$_3$ ($001$) peak with pronounced Laue-oscillations at an oxygen partial pressure of $6.8\cdot10^{-4}$\,mbar.}%
\label{o2partialdruckserie}%
\end{figure}%
\begin{figure}[!t]%
\centering%
\includegraphics[width=\linewidth]{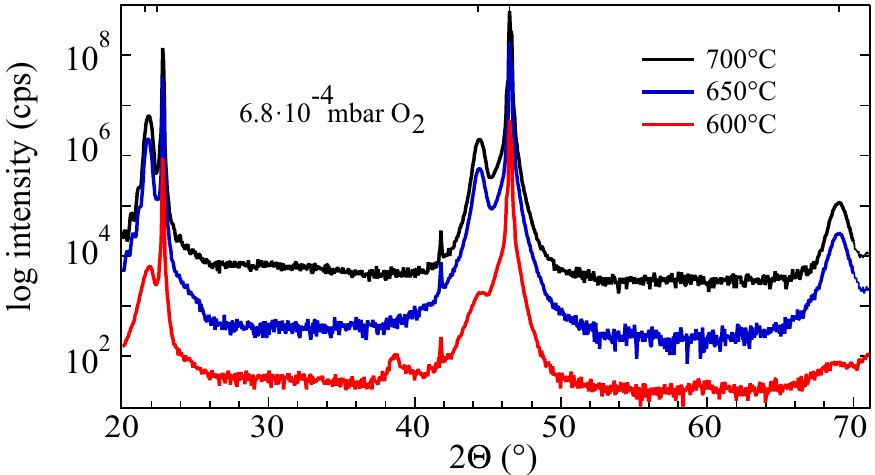}%
\caption{XRD patterns for BiFeO$_3$ films grown at a various substrate temperatures and an O$_2$ partial pressure of $6.8\cdot10^{-4}$\,mbar.}%
\label{temperaturserie}%
\end{figure}%
X-ray diffraction (XRD) and x-ray reflectivity (XRR) measurements were performed to obtain information about the crystallographic properties of the films. In order to determine the optimal growth conditions, the substrate temperature was set to $700^\circ$C and the oxygen partial pressure was varied from $5.3\cdot10^{-4}$\,mbar to $1.1\cdot10^{-3}$\,mbar. The Ar partial pressure was $1.5\cdot10^{-3}$\,mbar and remained constant for all samples. The corresponding XRD patterns are presented in Fig. \ref{o2partialdruckserie}. Films deposited with a small oxygen partial pressure of $5.3\cdot10^{-4}$\,mbar show a reflection around $2\Theta=39^\circ$, that may be caused by a small amount of pure Bi. At oxygen partial pressures between $6.8\cdot10^{-4}$\,mbar and $8.1\cdot10^{-4}$\,mbar no secondary phases are visible in the XRD patterns. At an oxygen partial pressure of $1.1\cdot10^{-3}$\,mbar the compound seems to be over-oxidized and the XRD pattern shows additional features (marked by arrows). The BiFeO$_3$ film grown at $6.8\cdot10^{-4}$\,mbar O$_2$ exhibits the sharpest peaks and pronounced Laue-oscillations, as shown in the inset in Fig. \ref{o2partialdruckserie}, and an out-of-plane lattice parameter of $4.08$\,\AA{}. Supposing an epitaxial matching of the in-plane BiFeO$_3$ lattice parameter to the SrTiO$_3$ substrate and assuming the volume of the BiFeO$_3$ unit cell remains constant upon strain, the expected lattice parameter is $c=4.05$\,\AA{}, which is very close to the measured one. To optimize the deposition temperature, the oxygen partial pressure of $6.8\cdot10^{-4}$\,mbar was kept constant and the substrate temperature was varied. The resulting XRD patterns are presented in Fig. \ref{temperaturserie}. The crystallization of BiFeO$_3$ starts at $600^\circ$C, resulting in small BiFeO$_3$ peaks. Again, the reflection at $2\Theta=39^\circ$ can be found. The lattice parameter decreases with higher deposition temperatures and comes closer to the expected values. At a deposition temperature of $750^\circ$C the BiFeO$_3$ film starts to evaporate as is indicated by greatly reduced film thickness for otherwise unchanged deposition conditions.

Consequently, an O$_2$ partial pressure of $6.8\cdot10^{-4}$\,mbar and a deposition temperature of $700^\circ$C are chosen as an ideal growth parameter for epitaxial BiFeO$_3$ films for further experiments on the exchange bias.

\section{X-ray absorption spectroscopy}
\begin{figure}[!t]%
\centering%
\includegraphics[width=\linewidth]{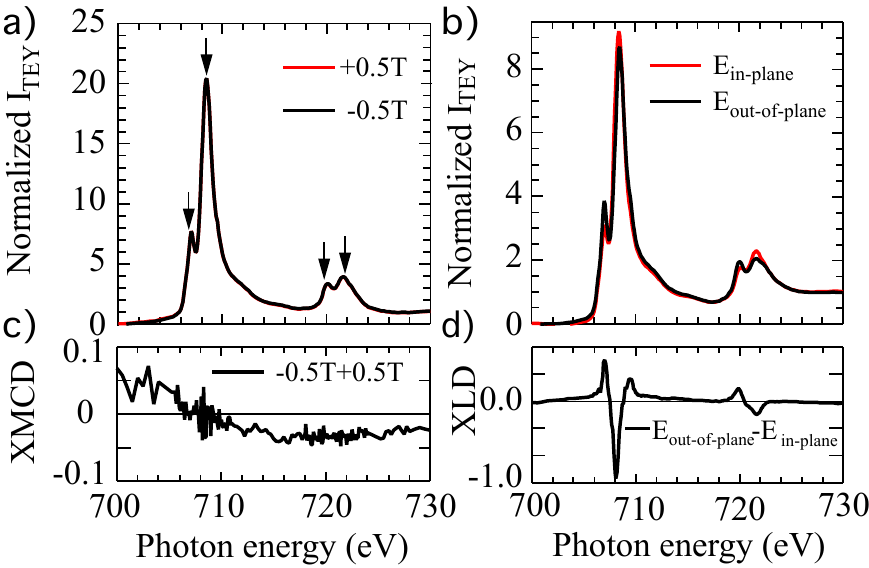}%
\caption{a) XAS measurements with both circular polarization with c) corresponding XMCD and b) XAS measurements with electric field component in- and out-of-plane with respect to the film together with the d) XLD difference.}%
\label{XMCD_XMLD_layout}%
\end{figure}%
Normalized XA spectra at the Fe L$_{2,3}$ absorption edge of optimized BiFeO$_3$ films are shown in Fig. \ref{XMCD_XMLD_layout} a). The two-peak structure is clearly visible at both the L$_2$- and L$_3$- edge, as expected for a trivalent Fe state (marked by the arrows) \cite{Bea:2006tf}. In pure BiFeO$_3$ the Fe atom is surrounded by oxygen atoms on a distorted octahedron. No difference is found between XA spectra with magnetization parallel or antiparallel to the beam, i.e., the XMCD is virtually zero as depicted in Fig. \ref{XMCD_XMLD_layout} c). Thus, our BiFeO$_3$ film shows no ferromagnetic order and contains no parasitic phases such as ferrimagnetic $\gamma$-Fe$_2$O$_3$, which is not easily detectable in XRD measurements in thin films \cite{Bea:2006tf,hunt}. To test the films for antiferromagnetism two XAS scans with the beam polarization vector parallel or perpendicular to the film plane were performed according to \cite{Kuiper:1993vg} and \cite{Bea:2006el}. The results are presented in Fig. \ref{XMCD_XMLD_layout} b). A clear difference between both orientations is visible as shown in Fig. \ref{XMCD_XMLD_layout} d). The XLD can be of structural or magnetic origin and these two possibilities cannot be distinguished from our measurements. However, our XLD spectra resemble those measured by B\'{e}a \textit{et al.} \cite{Bea:2006el} very closely in shape and signal amplitude. The antiferromagnetic order in their BiFeO$_3$ films was confirmed by neutron diffraction experiments, so we conclude that our BiFeO$_3$ films are most probably antiferromagnetically ordered at room temperature.

\section{Magnetic Properties}
\begin{figure}[!t]%
\centering%
\includegraphics[width=\linewidth]{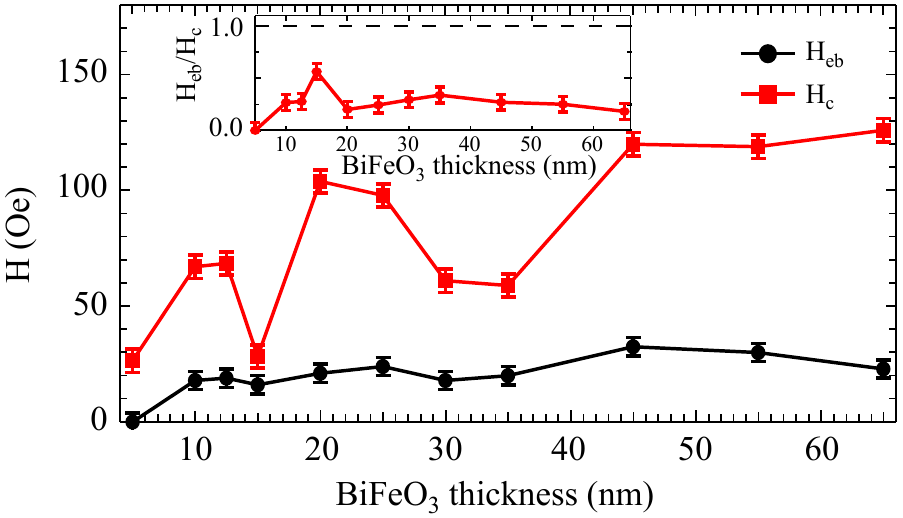}%
\caption{Exchange bias as a function of BiFeO$_3$ film thickness with a $3$\,nm Co$_{60}$Fe$_{40}$ layer. The H$_{\mathrm{eb}}$/H$_{\mathrm{c}}$ ratio is presented in the inset.}%
\label{H_eb_vs_BFO_thickness_Co6Fe40_L}%
\end{figure}%
\begin{figure}[!t]%
\centering%
\includegraphics[width=\linewidth]{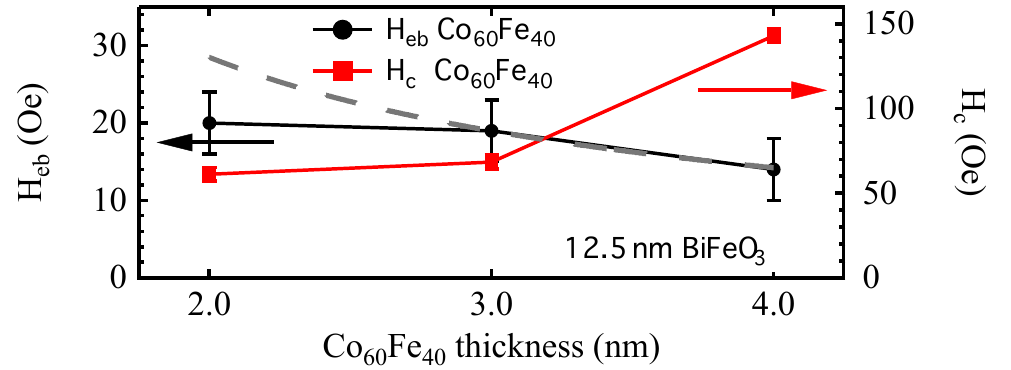}%
\caption{Echange bias for a $12.5$\,nm BiFeO$_3$ film with different Co$_{60}$Fe$_{40}$ film thicknesses.}%
\label{H_eb_vs_Co60Fe40_thickness_layo}%
\end{figure}
\begin{figure}[!t]%
\centering%
\includegraphics[width=\linewidth]{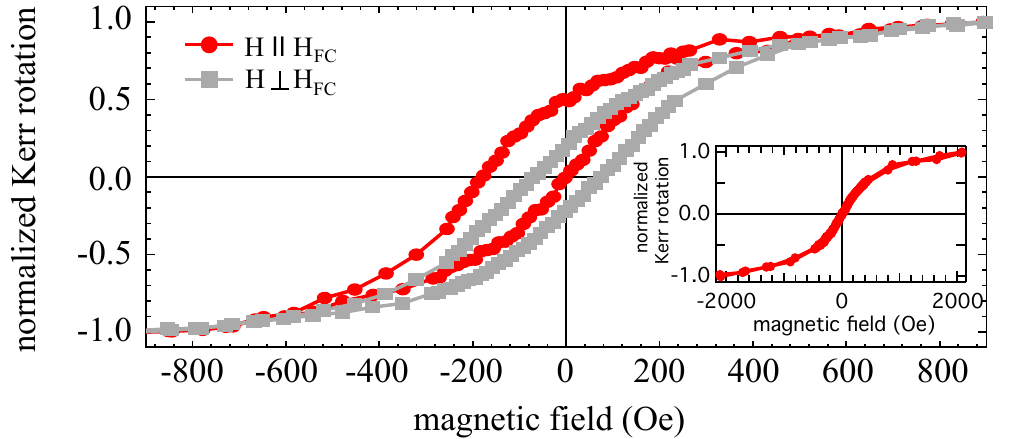}%
\caption{Hysteresis for a $12.5$\,nm BiFeO$_3$ / $2$\,nm Co film for parallel and perpendicular alignment with respect to the field cooling direction. The inset shows a MOKE measurement for a $12.5$\,nm BiFeO$_3$ / $1.5$\,nm Co bilayer.}%
\label{CS150613b_Grad_250_Hysterese}%
\end{figure}

%
The influence of the BiFeO$_3$ layer thickness on the exchange bias with a $3$\,nm thick Co$_{60}$Fe$_{40}$ film is presented in Fig. \ref{H_eb_vs_BFO_thickness_Co6Fe40_L}. The shift of the hysteresis H$_{\mathrm{eb}}$ as well as the coercive field H$_{\mathrm{c}}$ are determined by room-temperature MOKE measurements. We observe a positive exchange bias where the hysteresis is shifted toward the negative field cooling direction. One can see a clear correlation between H$_{\mathrm{eb}}$ and H$_{\mathrm{c}}$, as seen in the H$_{\mathrm{eb}}$/H$_{\mathrm{c}}$ ratio in the inset of Fig. \ref{H_eb_vs_BFO_thickness_Co6Fe40_L}. Both fields generally increase with increasing film thickness.  A maximum exchange bias field H$_{\mathrm{eb}}=33$\,Oe was found for $45$\,nm BiFeO$_3$. The apparently non-systematic behavior of the coercive field may be related to problems with the heat transfer between the substrate and the holder, that in some cases led to somewhat lower substrate temperatures.  A thickness as low as $12.5$\,nm BiFeO$_3$ generates an exchange bias of H$_{\mathrm{eb}}=19$\,Oe and a coercive field of H$_{\mathrm{c}}=69$\,Oe. This thickness is used to study the influence of different Co-Fe film thicknesses and Co-Fe stoichiometries. Fig. \ref{H_eb_vs_Co60Fe40_thickness_layo} shows H$_{\mathrm{eb}}$ and H$_{\mathrm{c}}$ as a function of Co$_{60}$Fe$_{40}$ thickness. A maximum exchange bias was found for a $2$\,nm thick Co$_{60}$Fe$_{40}$ film with H$_{\mathrm{eb}}=20$\,Oe and H$_{\mathrm{c}}=61$\,Oe. The coercivity increases with increasing ferromagnet thickness up to H$_{\mathrm{c}}=143$\,Oe at a thickness of $4$\,nm Co$_{60}$Fe$_{40}$. The usually expected proportionality H$_{\mathrm{eb}} \propto 1/t_{\mathrm{FM}}$ (indicated by the dashed line) cannot be found \cite{Nogues:1999tn}. A similar behavior is found in MnN/CoFe bilayers in reference \cite{Meinert:2015}. Co$_{60}$Fe$_{40}$ films thinner than $2$\,nm show no hysteresis with vanishing coercivity and no shift of the magnetization curve. This is probably responsible for the flattening of the H$_{\mathrm{eb}}$ curve around $2$\,nm. The inset in Fig. \ref{CS150613b_Grad_250_Hysterese} shows a MOKE measurement of a $12.5$\,nm BiFeO$_3$ / $1.5$\,nm Co bilayer.

The exchange bias can be increased by using ferromagnetic layers with a higher amount of Co and reaches its maximum for a pure Co film with H$_{\mathrm{eb}}=92$\,Oe and H$_{\mathrm{c}}=89$\,Oe. Fig. \ref{CS150613b_Grad_250_Hysterese} exemplarily shows the MOKE measurements for a $12.5$\,nm BiFeO$_3$ / $2$\,nm Co film. Moreover, no training effect was observed after $100$ loops \cite{Nogues:1999tn}.
\begin{table}
\centering%
\caption{\label{Max_exchange_bias_fields}Maximum exchange bias fields and coercive fields H$_{\mathrm{eb}}$ and H$_{\mathrm{c}}$ for various stoichiometries of a $2$\,nm thick ferromagnet for a $12.5$\,nm BiFeO$_3$ film, determined from MOKE measurements.}%
\begin{tabular}{|c|c|c|}%
   \hline  composition &  H$_{\mathrm{eb}}$ (Oe) & H$_{\mathrm{c}}$ (Oe)  \\%
   \hline  Fe & - & - \\%
   \hline  Co$_{60}$Fe$_{40}$  & $20$ & $61$ \\%
   \hline  Co$_{90}$Fe$_{10}$  & $36$ & $27$ \\%
   \hline  Co & $92$ & $89$\\ %
   \hline  %
\end{tabular}%
\end{table}%
Table \ref{Max_exchange_bias_fields} summarizes the maximum exchange bias and coercive fields observed for different compositions of the ferromagnetic layer with a thickness of $2$\,nm. For pure Fe films no ferromagnetic hysteresis can be observed, likely due to oxidation of the Fe film.
Within the Meiklejohn-Bean model the exchange energy \cite{Nogues:1999tn} is $J_{\mathrm{eff}}=t_{\mathrm{FM}}M_{\mathrm{FM}}H_{\mathrm{eb}}\approx 0.02 \frac{\mathrm{erg}}{\mathrm{cm}^2}$. Together with a critical BiFeO$_3$ film thickness of $t_{\mathrm{crit}} \approx 10$\,nm the effective unidirectional anisotropy constant is $K_{\mathrm{eff}}=J_{\mathrm{eff}}/t_{\mathrm{crit}}\approx2\cdot10^{4} \frac{\mathrm{erg}}{\mathrm{cm}^3}$. These values are much smaller than, for example, for Mn$_3$Ir with $J_{\mathrm{eff}}>1 \frac{\mathrm{erg}}{\mathrm{cm}^2}$ and $K_{\mathrm{eff}}=2\cdot10^{6}\frac{\mathrm{erg}}{\mathrm{cm}^3}$ \cite{Carey:2001ej}.

B\'{e}a \textit{et al.} \cite{Bea:2006el} found slightly smaller exchange bias fields for single crystalline $35$\,nm BiFeO$_3$ films with $5$\,nm Co$_{72}$Fe$_8$B$_{20}$ of H$_{\mathrm{c}}=42$\,Oe and H$_{\mathrm{eb}}=62$\,Oe. Measurements of a $200$\,nm BiFeO$_3$ / 5\,nm Ni$_{81}$Fe$_{19}$ bilayer by Dho \textit{et al.} \cite{Dho:2009da} revealed an exchange bias in the same order of magnitude of H$_{\mathrm{eb}}=80$\,Oe but with a lower coercivity of H$_{\mathrm{c}}=22$\,Oe.
Polycrystalline BiFeO$_3$ films are reported to generate a much larger exchange bias effect than single crystalline films. Chen \textit{et al.} \cite{Chang:2012ed} found exchange bias fields of up to H$_{\mathrm{eb}}=400$\,Oe together with coercive fields of H$_{\mathrm{c}}=1201-3632$\,Oe.

\begin{figure}[!t]%
\centering%
\includegraphics[width=\linewidth]{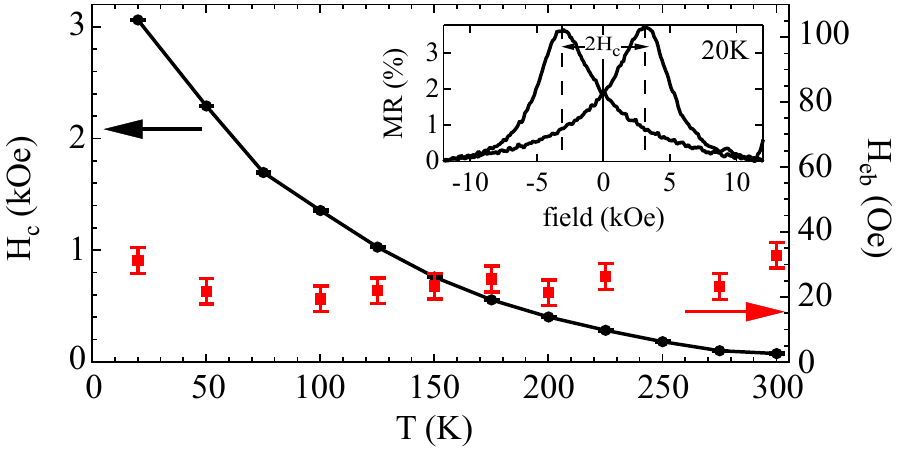}%
\caption{Coercive field H$_{\mathrm{c}}$ and exchange bias field H$_{\mathrm{eb}}$ versus measurement temperature for a $12.5$\,nm BiFeO$_3$/ $2$\,nm Co bilayer obtained by AMR measurements. The inset shows exemplarily the AMR measurement at $20$\,K. The MR is normalized by $MR=(R(H)-R_{\mathrm{min}})/R_{\mathrm{min}}\cdot 100\%$.}%
\label{Hc_vs_temp}%
\end{figure}

The temperature dependence of the exchange bias of a $12.5$\,nm BiFeO$_3$/ $2$\,nm Co sample was determined by AMR measurements. The field cooling direction was aligned parallel to the applied magnetic field and as well as the probe current. The inset in Fig. \ref{Hc_vs_temp} exemplarily shows the AMR curve measured at $20$\,K. The peak positions were fitted using two Gaussian functions to determine the coercive field H$_{\mathrm{c}}$ as well as the exchange bias field H$_{\mathrm{eb}}$. The results are presented in Fig. \ref{Hc_vs_temp}. The coercivity increases from H$_{\mathrm{c}}=78$\,Oe at $300$\,K to H$_{\mathrm{c}}=3.1$\,kOe at $20$\,K. This behavior is very similar to the temperature dependence measured in NiFe/PtPdMn and NiFe/NiMn systems by Hou \textit{et al.} \cite{PhysRevB.63.024411}. A shift in the AMR curves determines the exchange bias to approximately H$_{\mathrm{eb}}=20$\,Oe, which remains constant for all temperatures. Rotation of the sample by $180^\circ$ degrees results in a corresponding shift to negative values and H$_{\mathrm{eb}}=-20$\,Oe.
\subsection{Anisotropy}
\begin{figure}[!t]%
\centering%
\includegraphics[width=\linewidth]{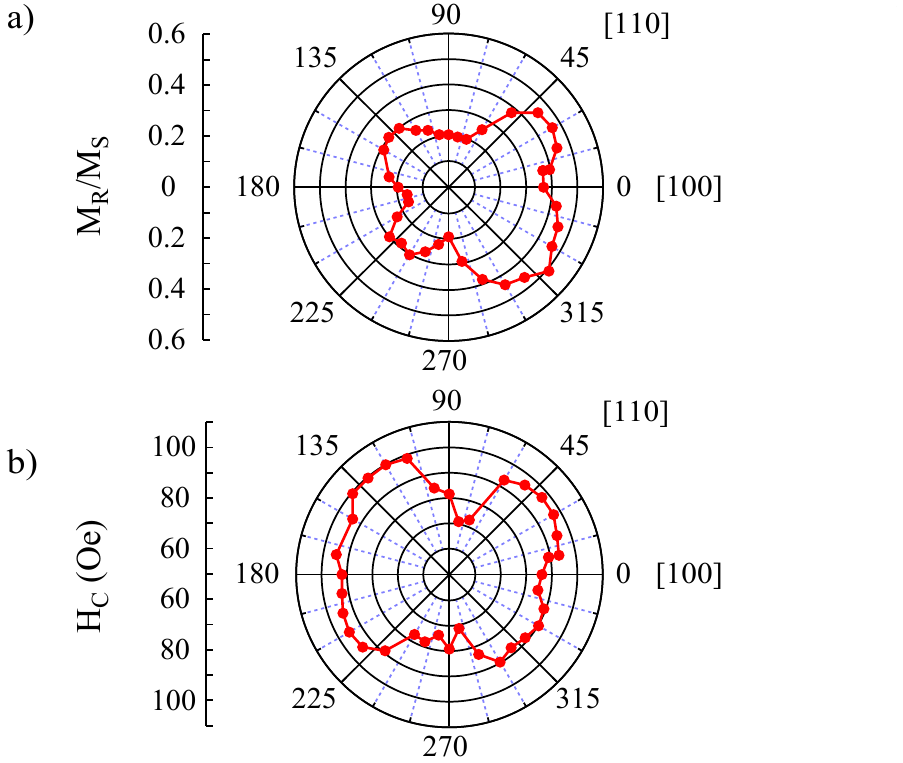}%
\caption{MOKE a) squareness M$_{\mathrm{R}}$/M$_{\mathrm{S}}$ and b) coercive field H$_{\mathrm{c}}$ for $12.5$\,nm BiFeO$_3$/ $2$\,nm Co as a function of the in-plane rotational angle. The magnetic field during field-cooling was applied along the [$100$] direction.}%
\label{Polar_graphs}%
\end{figure}

As an additional confirmation of the presence of exchange bias in our films, in-plane MOKE rotation measurements were performed. The squareness M$_{\mathrm{R}}$/M$_{\mathrm{S}}$ and coercive field H$_\mathrm{c}$ versus rotational angle are presented in Fig. \ref{Polar_graphs} a) and b). The field-cooling direction is in [$100$] direction of the BiFeO$_3$/Co films, which is defined as $0^\circ$.
The squareness shows a pronounced unidirectional anisotropy with a superimposed fourfold anisotropy. This indicates that the BiFeO$_3$ film induces a bct structure in the Co films \cite{Pires:2011kn}. The corresponding coercive field versus rotational angle (see Fig. \ref{Polar_graphs} b)) shows the butterfly-like shape typical of exchange bias system with a weak superimposed fourfold anisotropy. As expected, the coercive field is small for angles perpendicular to the field cooling direction, i.e. in the [$010$] and [$0\bar{1}0$] directions.

\section{Conclusion}
In summary, we have found a maximum exchange bias of H$_{\mathrm{eb}}=92$\,Oe and H$_{\mathrm{c}}=89$\,Oe for a $12.5$\,nm BiFeO$_3$/ $2$\,nm Co bilayer. At low temperatures no increase in the exchange bias was found. The coercivity of the pinned Co layer increases of up to $3$\,kOe at $20$\,K. MOKE rotational measurements revealed a fourfold anisotropy of the exchange-biased Co layer together with a pronounced unidirectional anisotropy component.

\section*{Acknowledgments}
The authors gratefully acknowledge financial support from Deutsche Forschungsgemeinschaft (DFG, contract No. RE 1052/32-1).


\begin{thebibliography}{10}
\providecommand{\url}[1]{#1}
\csname url@samestyle\endcsname
\providecommand{\newblock}{\relax}
\providecommand{\bibinfo}[2]{#2}
\providecommand{\BIBentrySTDinterwordspacing}{\spaceskip=0pt\relax}
\providecommand{\BIBentryALTinterwordstretchfactor}{4}
\providecommand{\BIBentryALTinterwordspacing}{\spaceskip=\fontdimen2\font plus
\BIBentryALTinterwordstretchfactor\fontdimen3\font minus
  \fontdimen4\font\relax}
\providecommand{\BIBforeignlanguage}[2]{{%
\expandafter\ifx\csname l@#1\endcsname\relax
\typeout{** WARNING: IEEEtran.bst: No hyphenation pattern has been}%
\typeout{** loaded for the language `#1'. Using the pattern for}%
\typeout{** the default language instead.}%
\else
\language=\csname l@#1\endcsname
\fi
#2}}
\providecommand{\BIBdecl}{\relax}
\BIBdecl


\bibitem{Eerenstein:2006km}
W.~Eerenstein, N.~D. Mathur, and J.~F. Scott, ``{Multiferroic and
  magnetoelectric materials}'', \emph{Nature}, vol. 442, no. 7104, p. 759,
  2006.

\bibitem{Kiselev}
S.~V. Kiselev, R.~P. Ozerov, and G.~S. Zhdanov, ``{Detection of Magnetic Order
  in Ferroelectric BiFeO$_{3}$ by Neutron Diffraction}'', \emph{Soviet Physics
  Doklady}, vol.~7, p. 742, 1963.

\bibitem{Teague:1970tk}
J.~R. Teague, R.~Gerson, and W.~J. James, ``{Dielectric hysteresis in single
  crystal BiFeO$_{3}$}'', \emph{Solid state communications}, vol.~8, no.~13, p.
  1073, 1970.

\bibitem{Wang:2003ca}
J.~Wang, J.~B. Neaton, H.~Zheng, V.~Nagarajan, S.~B. Ogale, B.~Liu,
  D.~Viehland, V.~Vaithyanathan, D.~G. Schlom, and U.~V. Waghmare, ``{Epitaxial
  BiFeO$_{3}$ multiferroic thin film heterostructures}'', \emph{Science}, vol.
  299, no. 5613, p. 1719, 2003.

\bibitem{Idzerda:1994vq}
Y.~U. Idzerda, C.~T. Chen, H.~J. Lin, G.~Meigs, and G.~H. Ho, ``{Soft X-ray
  magnetic circular dichroism and magnetic films}'', \emph{Nuclear Instruments
  and Methods in Physics Research Section A: Accelerators, Spectrometers,
  Detectors and Associated Equipment}, vol. 347, no. 1-3, p. 134, 1994.

\bibitem{Bea:2006tf}
H.~B{\'e}a, M.~Bibes, S.~Fusil, K.~Bouzehouane, E.~Jacquet, K.~Rode, P.~Bencok,
  and A.~Barth{\'e}l{\'e}my, ``{Investigation on the origin of the magnetic
  moment of BiFeO$_{3}$ thin films by advanced x-ray characterizations}'',
  \emph{Physical Review B}, vol.~74, no.~2, p. 020101, 2006.

\bibitem{hunt}
C.~P. Hunt, B.~M. Moskowitz, and S.~K. Banerjee, ``{Magnetic Properties of
  Rocks and Minerals. Rock Physics {\&} Phase Relations: A Handbook of Physical
  Constants}'', \emph{American Geophysical Union, Washington DC}, vol.~3, p.
  189, Jan. 1995.

\bibitem{Kuiper:1993vg}
P.~Kuiper, B.~G. Searle, P.~Rudolf, L.~H. Tjeng, and C.~T. Chen, ``{X-ray
  magnetic dichroism of antiferromagnet Fe$_{2}$O$_{3}$: The orientation of
  magnetic moments observed by Fe 2p X-ray absorption spectroscopy}'',
  \emph{Physical Review Letters}, vol.~70, no.~10, p. 1549, 1993.

\bibitem{Bea:2006el}
H.~B{\'e}a, M.~Bibes, S.~Cherifi, F.~Nolting, B.~Warot-Fonrose, S.~Fusil,
  G.~Herranz, C.~Deranlot, E.~Jacquet, and K.~Bouzehouane, ``{Tunnel
  magnetoresistance and robust room temperature exchange bias with multiferroic
  BiFeO$_{3}$ epitaxial thin films}'', \emph{Applied Physics Letters}, vol.~89,
  p. 242114, 2006.

\bibitem{Nogues:1999tn}
J.~Nogu{\'e}s and I.~K. Schuller, ``Exchange bias'', \emph{Journal of Magnetism
  and Magnetic Materials}, vol. 192, no.~2, p. 203, 1999.

\bibitem{Meinert:2015}
M.~Meinert, B.~B\"uker, D.~Graulich, and M.~Dunz, ``{Large exchange bias in
  polycrystalline MnN/CoFe bilayers at room temperature}'', \emph{Phys. Rev.
  B}, vol.~92, p. 144408, 2015.

\bibitem{Carey:2001ej}
M.~J. Carey, N.~Smith, and B.~A. Gurney, ``{Thermally assisted decay of pinning
  in polycrystalline exchange biased systems}'', \emph{Journal of Applied
  Physics}, vol.~89, no.~11, p. 6579, 2001.

\bibitem{Dho:2009da}
J.~Dho and M.~G. Blamire, ``{Controlling the exchange bias in multiferroic
  BiFeO$_{3}$ and NiFe bilayers}'', \emph{Journal of Applied Physics}, vol.
  106, no.~7, p. 073914, 2009.

\bibitem{Chang:2012ed}
H.~W. Chang, F.~T. Yuan, C.~W. Shih, W.~L. Li, P.~H. Chen, C.~R. Wang, W.~C.
  Chang, and S.~U. Jen, ``{Exchange bias in sputtered FM/BiFeO$_{3}$ thin films
  (FM = Fe and Co)}'', \emph{Journal of Applied Physics}, vol. 111, no.~7, p.
  07B105, 2012.

\bibitem{PhysRevB.63.024411}
C.~Hou, H.~Fujiwara, K.~Zhang, A.~Tanaka, and Y.~Shimizu, ``Temperature
  dependence of the exchange-bias field of ferromagnetic layers coupled with
  antiferromagnetic layers'', \emph{Phys. Rev. B}, vol.~63, p. 024411, 2000.

\bibitem{Pires:2011kn}
M.~Pires, A.~Cotta, M.~D. Martins, and A.~Silva, ``{Four-fold magnetic
  anisotropy in a Co film on MgO (001)}'', \emph{Journal of Magnetism and
  Magnetic Materials}, vol. 323, no.~6, p. 789, 2011.

\end{thebibliography}
\end{document}